\documentclass[aps,pre,twocolumn,a4]{revtex4}
\usepackage{times} 
\usepackage{amssymb}
\usepackage{amsbsy}
\usepackage{amsmath}
\usepackage{color}
\usepackage{ulem}

\usepackage{prelim2e} 
\usepackage[none,bottom]{draftcopy}
\draftcopyName{UT Preprint / }{1.2}

\ifx\pdfoutput\undefined
\usepackage{graphicx}
\else
\usepackage[pdftex]{graphicx,hyperref}
\fi
\graphicspath{{.},{../../Fig/},,{../../Ps/}{../../Figs_nottingham/},{$LOCALHOME/Obj/Sitdrop/Ps/},%
{$LOCALHOME/Obj/Nanopart/Ps/},{$LOCALHOME/Obj/Nanocont/Fig/},
{$LOCALHOME/Literatur/Ps/},{$LOCALHOME/Obj/DDFT/Fig/}}
\newcommand {\dr}{{\mathrm d}\mathbf{r}}

\newcommand {\rr}{\mathbf{r}}

\begin{document}
\title{Dynamical density functional theory for the dewetting of evaporating thin films of nanoparticle suspensions exhibiting pattern formation}
\author{A.J.~Archer} 
\author{M.J.~Robbins}
\author{U.~Thiele}
\affiliation{Department of Mathematical Sciences, 
Loughborough University, Leicestershire LE11 3TU, UK}

\begin{abstract}
  Recent experiments have shown that the striking structure formation in 
 dewetting films of evaporating colloidal nanoparticle suspensions 
  occurs in an ultrathin `postcursor' layer that is left behind by a mesoscopic
  dewetting front. Various phase change and transport processes occur in
  the postcursor layer, that may lead to nanoparticle deposits in the form of
  labyrinthine, network or strongly branched `finger' structures. We develop a
  versatile dynamical density functional theory
  to model this system which captures all these structures and may be employed
  to investigate the influence of evaporation/condensation,
  nanoparticle transport and solute transport in a differentiated way.
  We highlight, in particular, the influence of the subtle
  interplay of decomposition in the layer and contact line motion on
  the observed particle-induced transverse instability of the
  dewetting front.
\end{abstract}
\maketitle

How surface patterns and structures evolve over time is of
great interest for a wide range of scientific fields. Striking
examples include river network patterns \cite{GMB95}, the
growth of rocks around geothermal springs \cite{VeGo08}
evaporation-caused coffee stain patterns \cite{Deeg00}, and the
patterns in the distribution of living organisms \cite{MPV08}. Many such
structures are generated by the interaction of fluid motion over the
surface and deposition and/or abrasion of material. A particular
process of high recent interest that concerns us in this Letter
is the formation of structures during the (evaporative) dewetting of
nanoparticle suspensions on solid substrates
\cite{GeBr00,MTB02,Paul08}. The patterning is generic
to a wide class of dewetting evaporating suspensions and solutions
\cite{GRDK02,LZZZ08,TMP98,Smal06,ZMCZ08} and depends crucially on the
interplay of several competing phase-change and transport
processes.

The rapidly expanding study of such systems currently receives strong
impetus from research in two distinct areas. On the one hand, studies
from the last couple of decades of the dynamics of dewetting of
surfaces by non-volatile liquids \cite{Reit92,SHJ01} have been
extended to investigate the interplay between (de)wetting and
evaporation of volatile liquids \cite{ElLi94,GIMK06}. On the other
hand, there is interest in the non-equilibrium thermodynamics and
rheology of the respective phase and flow behaviour of bulk
suspensions and solutions -- see e.g.\ Ref.~\cite{QuBe02} and
references therein. A thin film of pure non-volatile liquid
that is deposited upon a smooth substrate (e.g., a polystyrene film
having a thickness of a few tens of nanometers, deposited on silicon
oxide \cite{Reit92}) may rupture due to effective molecular interactions
between the film surface and the solid substrate. The rupture mechanism
can be (i) via a surface instability (often
called spinodal dewetting) that occurs spontaneously and results
in patterns of a certain characteristic wavelength, or (ii) via
nucleation at randomly distributed defects \cite{TVN01,Beck03}.
The resulting holes then grow to form a polygonal network of liquid
rims that may subsequently decay into drops. All stages of this
process are intensively studied: the rupture mechanisms 
\cite{Reit92,SHJ01,TVN01}, the hole growth \cite{RBR91}, the morphologies and
evolution of the resulting patterns \cite{ShRe96}, and the stability
of receding liquid fronts \cite{ReSh01,MuWa05}. For reviews of this
body of work, see Ref.\ \cite{KaTh07}.

The dewetting processes of solutions and suspensions are more involved
than those of a pure liquid because they involve several
interdependent dynamical processes: transport of solute or colloids,
transport of the solvent and evaporation/condensation of the solvent.
As a consequence, one must distinguish between `normal' convective
dewetting and {\it evaporative} dewetting.  Experimental studies
performed with volatile solutions/suspensions of polymers
\cite{GRDK02,LZZZ08}, macromolecules \cite{TMP98,Smal06,ZMCZ08} and
colloids a few nanometers in size (referred to as `nanoparticles')
\cite{MMNP00,GeBr00,MTB02,Paul08} describe a variety of richly
structured deposits of the solutes. One may observe labyrinthine and
polygonal network structures similar to the structures observed
following `classical' dewetting. As the solvent evaporates, the solute
remains dried onto the substrate and therefore `conserves' the transient
dewetting pattern \cite{TMP98,MTB02}. However, the solute is
not just a passive tracer: it may influence the thresholds and the
rates of the initial film rupture processes. Most importantly, it
may also destabilise the straight dewetting fronts and trigger the
creation of strongly ramified structures -- a process observed in
many different systems \cite{Thie98,GRDK02,Paul08,LZZZ08}.

For instance, the dewetting of films of a suspension of thiol-coated
gold nanoparticles in toluene may result in the deposition of the
nanoparticles in branched finger patterns. The precise properties of
these depend on the strength of the attraction between the colloidal
particles \cite{Paul08}. Employing contrast-enhanced video microscopy
to study the dynamics of the system, initially the receding of
a mesoscopic dewetting front (equivalent to the receding three phase
contact line) is observed, leaving behind an unstable ultrathin
`postcursor' film, having a thickness similar to the diameter of the
nanoparticles. Subsequently, the postcursor film breaks, forming a
pattern of holes that themselves grow in an unstable manner, resulting
in an array of branched structures. Note that the mesoscopic front may
also be unstable -- an effect that does not concern us here.

Theories for modelling such processes are rather limited at the
present time.  Hydrodynamical models of dewetting by thin films based
on a long-wave approximation \cite{ODB97,KaTh07} provide a mesoscale
description of the liquid film, and can account for evaporation of
volatile liquids \cite{LGP02} and for the presence of a solute
\cite{WCM03}. However, they are not able to describe the
processes occurring in the postcursor film as they do not account
for the interactions between the solute particles and between the
solute particles and the solvent. Alternatively, to describe these processes one
may employ two-dimensional (2d) kinetic Monte-Carlo
(KMC) lattice models that focus solely on the dynamics of the solute
diffusion and the solvent evaporation
\cite{RRGB03,Mart07,Vanc08}. The neglect of convective solvent
transport can be justified based on estimates comparing its
influence with the one of evaporative solvent transport
\cite{Vanc08}. However, so far the 2d KMC models have not
incorporated diffusive solvent transport that might be important in
the postcursor layer.

In this Letter we present an alternative description for the
structure formation in the postcursor film that does not have the
limitations of the previous approaches. We develop a 2d dynamical
density functional theory (DDFT) \cite{MaTa99,ArRa04,GPDM03,footnote1} to
describe the coupled dynamics of the density fields of the liquid
$\rho_l(\rr,t)$ and the nanoparticles $\rho_n(\rr,t)$.  In this
approach, diffusive liquid transport can be incorporated in a
straightforward manner enabling us to go beyond previous
2d KMC studies and examine its influence.
To construct the DDFT model, we (i) develop
via coarse-graining an approximation for the underlying
free energy functional of the system and (ii) form equations governing
the dynamics of the two density fields. These are able to account for
the non-conserved and conserved aspects of the dynamics, i.e., phase
change and diffusive transport processes, respectively.

\begin{figure}
\centering
\includegraphics[width=1.\hsize]{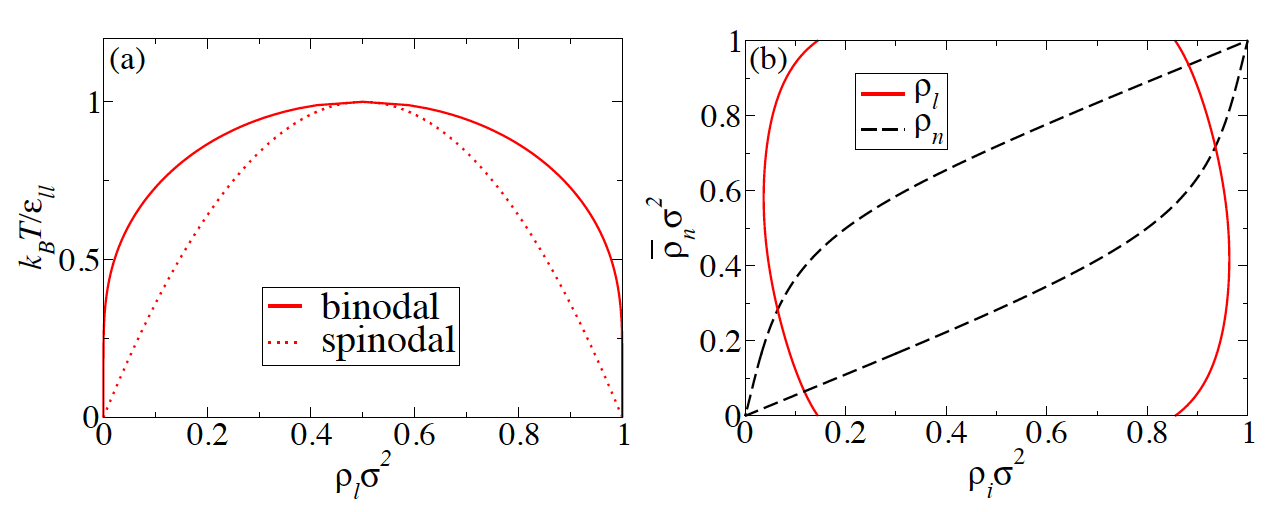}
\caption{(Color online) (a) Phase diagram of the pure liquid in the plane spanned
  by the density $\rho_l$ and temperature $T$. (b) Liquid and nanoparticle
  densities $\rho_l$ and $\rho_n$ at coexistence as a function of the average
  nanoparticle density $\bar{\rho}_n$, for $\varepsilon_{ll}/k_BT=1.25$,
  $\varepsilon_{nl}/k_BT=0.6$, $\varepsilon_{nn}=0$.}
   \label{fig:phase_diag}
\end{figure}

In order to compare results with the established KMC lattice model
\cite{RRGB03,Mart07, 
Vanc08}, we start from the lattice Hamiltonian
to develop a mean-field (Bragg-Williams) approximation
for the free energy functional \cite{ChLu97,GPDM03}. Expressing the
interaction terms (sums over neighbouring lattice sites)
as gradient operators \cite{GPDM03}, the following semi-grand
\cite{footnote} free energy functional is obtained:
\begin{eqnarray}
F[\rho_l,\rho_n]=\int \dr \bigg[ f(\rho_l,\rho_n)
+\frac{\varepsilon_{ll}}{2}(\nabla \rho_l)^2 
+\frac{\varepsilon_{nn}  }{2}(\nabla \rho_n)^2\nonumber \\
+\varepsilon_{nl}  (\nabla \rho_n)\cdot (\nabla \rho_l) 
-\mu\rho_l \bigg],
\label{eq:free_energy}
\end{eqnarray}
where $f(\rho_l,\rho_n)=k_BT[\rho_l\ln \rho_l+(1-\rho_l) \ln
(1-\rho_l) ] +k_BT [\rho_n \ln \rho_n+(1-\rho_n) \ln (1-\rho_n) ] -2
\varepsilon_{ll} \rho_l^2-2 \varepsilon_{nn} \rho_n^2
-4\varepsilon_{nl} \rho_n \rho_l,$ includes entropic contributions and
various interaction terms -- the parameters $\varepsilon_{ij}$, where
$i,j=n,l$, are the energies for having neighbouring pairs of lattice
sites occupied by species $i$ and $j$, respectively, $T$ is the
temperature, $k_B$ is Boltzmann's constant and we have set the lattice
spacing $\sigma=1$.  Note that $F$ in Eq.~(\ref{eq:free_energy}) can
also be obtained by making a gradient expansion of the (non-local)
free energy functional of a continuous system \cite{ChLu97}.
The rate of evaporation of the liquid from the substrate is determined
by the chemical potential $\mu$ in the reservoir (i.e.\ in the vapour
above the substrate).  When the temperature is sufficiently low, and
the chemical potential $\mu=\mu_{\rm coex}$, we observe coexistence
between a thick (high density) and a thin (low density) liquid
film. In Fig.\ 1(a) we display the limit of linear stability
(spinodal) and the equilibrium coexistence curve (binodal) for the
pure liquid (i.e.\ with $\rho_n=0$).  To indicate the influence of the
solute we plot in Fig.\ 1(b) the densities $\rho_l$ and $\rho_n$ at
coexistence for a fixed temperature $k_BT/\varepsilon_{ll}=0.8$ as a
function of the average concentration
$\bar{\rho}_n=\frac{1}{2}(\rho_n^a+\rho_n^b)$, where $\rho_n^a$ and
$\rho_n^b$ are the densities of the nanoparticles in the coexisting
$a$ and $b$ phases. For further details concerning the phase diagram
and its topology see Ref.\ \cite{WoSc06}.

\begin{figure}
\centering
\includegraphics[width=1.\hsize]{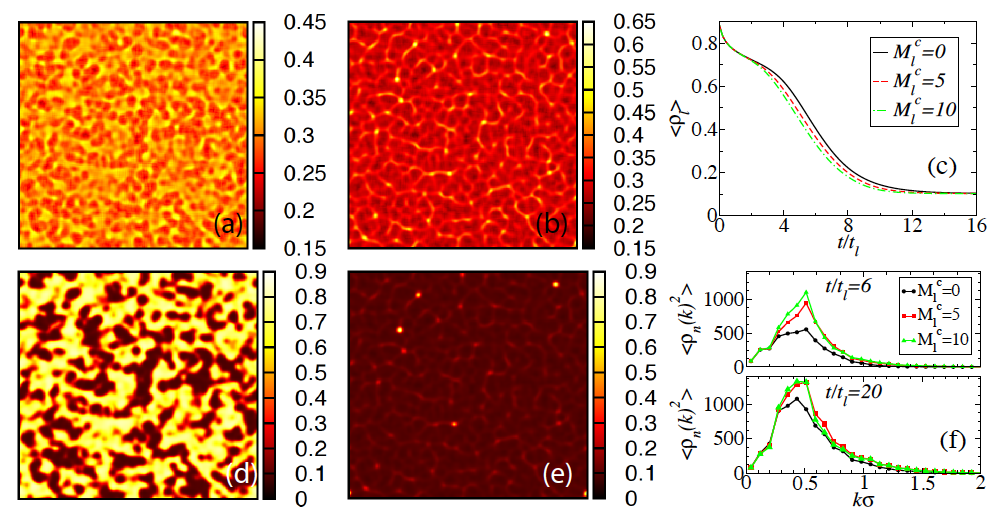}
\caption{(Color online) Results for spinodal evaporative dewetting of a
  nanoparticle suspension. (a) and (b) are typical nanoparticle
  density profiles, for times $t/t_l=6$ and $20$, and (d) and (e) are the corresponding
  liquid density profiles, for $M_l^c=0$.
  The domain size is $200\sigma \times 200\sigma$. (c) and (f)
  give the corresponding time evolution of the mean liquid density $\langle\rho_l\rangle$
  and the structure factor $\langle\rho_n(k)^2\rangle$ for various
  values of $M_l^c$.  The remaining
  parameters are $\varepsilon_{ll}/k_BT=1.25$,
  $\varepsilon_{nl}/k_BT=0.6$, $\varepsilon_{nn}=0$,
  $\alpha=0.4M_l^\mathrm{nc}\sigma^4$,
  $\mu/k_BT=-3.4$, and $\langle\rho_n\rangle=0.3$.
}
\label{fig:ddft_spinodal}
\end{figure}

At equilibrium, the derivative $\mu_n\equiv \delta F[\rho_n,\rho_l]/\delta
\rho_n$ is a constant, corresponding to the chemical potential of the
nanoparticles.
However, when the system is out of equilibrium, $\mu_n$ may vary
along the substrate.
We assume that the thermodynamic force $\nabla \mu_n$ drives the
dynamics of the nanoparticles and that
the nanoparticle current is ${\bf j}=-M_n \rho_n \nabla
\mu_n$, where $M_n(\rho_l)$ is a mobility coefficient. This expression for the
current, together with the continuity equation, yields
the time evolution equation for the nanoparticle density profile:
\begin{equation}
\frac{\partial \rho_n}{\partial t}=
\nabla \cdot \left[ M_n\rho_n \nabla \frac{\delta F[\rho_n,\rho_l]}{\delta \rho_n}\right].
\label{eq:DDFT_n}
\end{equation}
This equation may also be obtained by assuming over-damped
stochastic equations of motion for the nanoparticles
\cite{MaTa99,ArRa04}.
To model the fact that nanoparticles do not diffuse over the dry
substrate (when $\rho_l$ is small) we set the mobility
$M_n(\rho_l)$ to switch at $\rho_l=0.5$ (smoothly) from zero for the 
dry substrate (low $\rho_l$) to $\alpha$ for the wet substrate (high $\rho_l$). 
Note that our results are not sensitive to the precise
form of $M_n(\rho_l)$.

\begin{figure}
\centering
\includegraphics[width=1.\hsize]{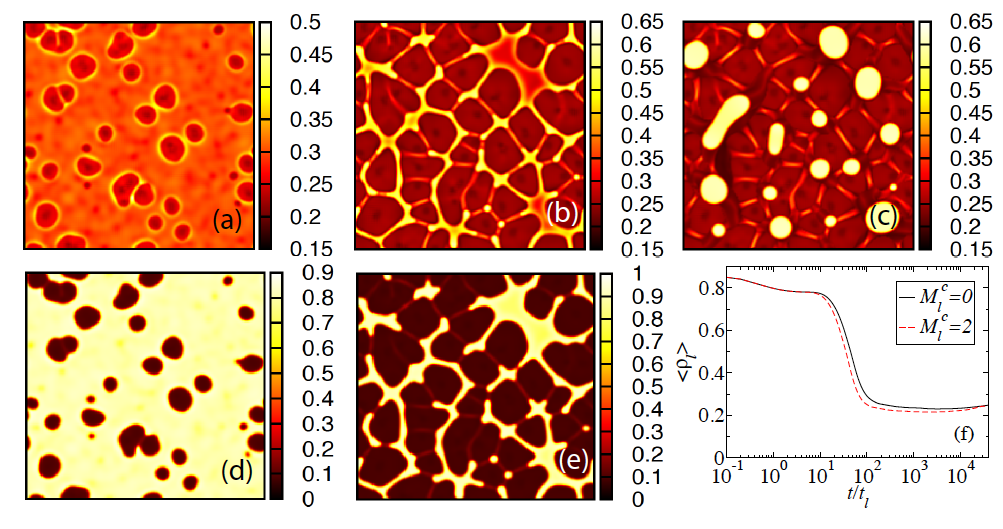}
\caption{(Color online) Density profiles for evaporative dewetting in the nucleation
  regime. (a)-(c) are the nanoparticle density
  profiles at times $t/t_l=20$, $80$ and $4000$, (d) and (e) are the liquid
  profiles for $t/t_l=20$ and $80$, for
  $M_l^c=2$ and $\mu/k_BT=-3.33$; remaining parameters are as in
  Fig.~\ref{fig:ddft_spinodal}. 
  In (f) we plot the average
  density of the liquid on the substrate $\langle \rho_l \rangle$, as a
  function of time for this case and the case $M_l^c=0$.  The system
  was initialised with the (discretised) density profiles:
  $\rho_l(x,y,t=0)=0.9+0.05\chi$, $\rho_n(x,y,t=0)=0.3+0.27\chi$, where
  $\chi$ is a random number uniformly distributed on the interval
  $[-1,1]$. It is due to this random noise that the holes are
  nucleated in some places and not in others.}
\label{fig:ddft_nucleation}
\end{figure}

\begin{figure}
\centering
\includegraphics[width=1.\hsize]{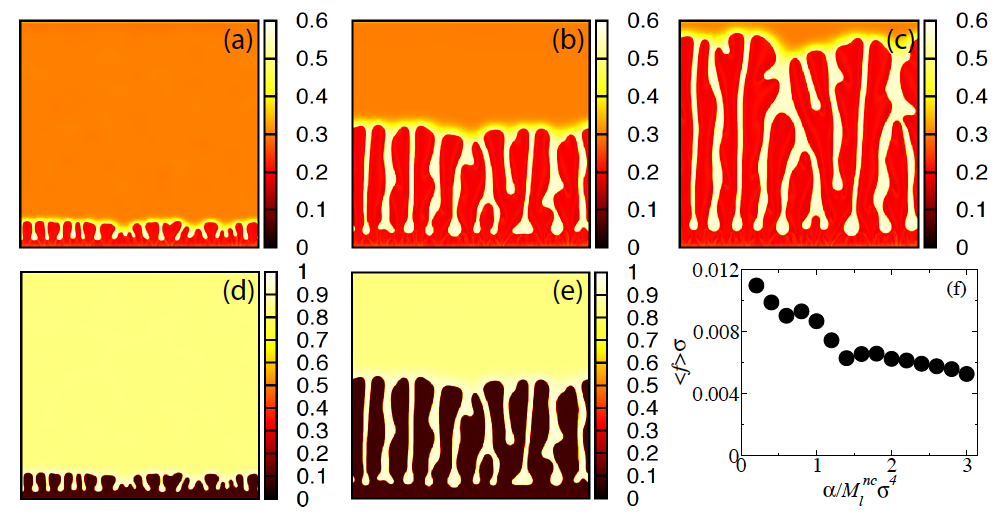}
\caption{(Color online) Density profiles from the evolution of an unstable dewetting front.
  (a)--(c) are the nanoparticle density profiles
  at times $t/t_l=2000$, $20000$ and $40000$, 
  (d) and (e) are the liquid profiles for $t/t_l=2000$ and $20000$,
  for $\alpha=0.2 M_l^\mathrm{nc}\sigma^4$,
  $\mu/k_BT=-3.28$ and domain size $800\sigma \times 800\sigma$;
  remaining parameters are as in Fig.~\ref{fig:ddft_spinodal}.
  The initial density profiles are the same as those in Fig.\ \ref{fig:ddft_nucleation},
  except for $y<0$ we set $\rho_l=\rho_n=10^{-10}$, to create an initially straight
  dewetting front at $y=0$. In (f) we plot
  the mean finger number $\langle f \rangle$ as a function of the mobility
  coefficient for nanoparticle diffusion $\alpha$, for the case when $M_l^c=0$. 
}
\label{fig:ddft_fingers}
\end{figure}

\begin{figure}[htb] 
\centering
\includegraphics[width=1.\hsize]{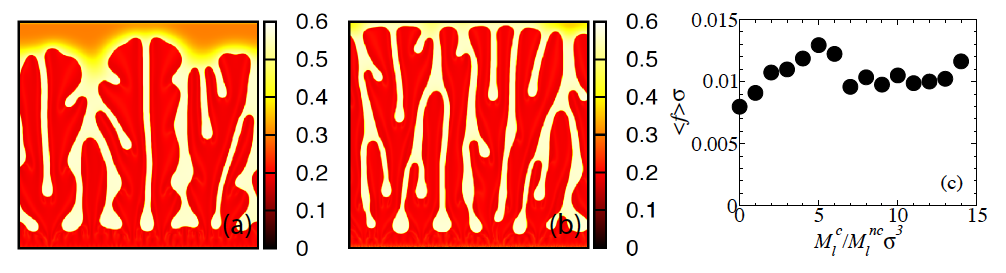}
\caption{(Color online) Nanoparticle density profiles from the evolution of an unstable
  dewetting front, for the time $t/t_l=20000$, (a) for $M_l^c=0$ and (b) $M_l^c=5$,
  for $\alpha=0.5 M_l^\mathrm{nc}\sigma^4$,
  $\mu/k_BT=-3.28$ and domain size is $800\sigma \times 800\sigma$;
  remaining parameters and initial density profiles are as in Fig.~\ref{fig:ddft_fingers}.
  In (c) we show the dependence
  of the mean finger number $\langle f \rangle$ on the mobility coefficient for
  liquid diffusion, $M_l^c$.}
\label{fig:fingers_vs_Mlc}
\end{figure}

For the liquid, the density may change either
by evaporation/condensation from/to the substrate (non-conserved dynamics)
or may diffuse over the substrate (conserved dynamics). The
latter dynamics is treated in a manner analogous to that for the nanoparticles.
For the non-conserved dynamics, we assume a standard
form \cite{Lang92}, i.e., that the change of the density over time is
proportional to $-(\mu_\mathrm{surf}-\mu)=-\delta F[\rho_n,\rho_l]/\delta
\rho_l$, where $\mu_\mathrm{surf}(\rr,t)$ is the local chemical
potential of the liquid on the substrate. Combining these two contributions,
we obtain the time evolution equation of the liquid density profile:
\begin{equation}
\frac{\partial \rho_l}{\partial t}=\nabla \cdot 
\left[ M_l^\mathrm{c}\rho_l \nabla \frac{\delta F[\rho_n,\rho_l]}{\delta \rho_l}\right]
- M_l^\mathrm{nc} \frac{\delta F[\rho_n,\rho_l]}{\delta \rho_l}.
\label{eq:DDFT_l}
\end{equation}
We assume that the two mobility coefficients
$M_l^\mathrm{c}$ and $M_l^\mathrm{nc}$ are constants.
In what follows we set $k_BT=1$ and $M_l^{nc}=1$.
Note that in the low density limit, when $\rho_n \to 0$ and $\rho_l\to 0$,
the conserved part in both Eq.\ \eqref{eq:DDFT_n} and
Eq.\ \eqref{eq:DDFT_l} corresponds to Fickian diffusion.

Before discussing results from our theory, we
first make a couple of comments about the status of the theory.
Firstly, we note that Eq.\ \eqref{eq:free_energy} constitutes a simple `zeroth-order'
mean-field approximation for the free energy of the system and omits
(for example) terms such as $\ln(1-\rho_n-\rho_l)$ which describe the
excluded area correlations between the liquid and the nano-particles.
Second, due to the fact that we derive the theory from the (already)
coarse-grained lattice Hamiltonian rather than by integrating over degrees
of freedom (coarse-graining) in the full DDFT theory for the three-dimensional
liquid film, the theory can not be regarded as a `fully' microscopic theory.
The reason that we have modelled the system using this simple theory
is because our interest is the basic question of what physics drives the
behaviour displayed in the experiments. In this work we choose to start
from the lattice theory, in order to compare with the KMC. However, as
future work we plan to go beyond the lattice theory. Our goal here is not to
construct a model describing every detail of these systems; instead
we seek to examine what physics is involved in determining the
observed pattern formation. Nonetheless, this DDFT does allow
us to investigate the time evolution of the
postcursor film of an evaporating nanoparticle suspension under fewer
restrictions than the KMC model. To discuss the importance of
liquid transport within the layer we compare results obtained for
different liquid mobilities ($M_l^c>0$) and results without liquid
transport (i.e., setting $M_l^c=0$). We focus on three examples: (i)
spinodal dewetting (see Fig.~\ref{fig:ddft_spinodal}), (ii) dewetting
via nucleation of holes in an initially flat film (see
Fig.~\ref{fig:ddft_nucleation}), and (iii) the unstable receding of an
evaporative dewetting front, which exhibits branched fingering (see
Figs.~\ref{fig:ddft_fingers} and \ref{fig:fingers_vs_Mlc}).

Fig.~\ref{fig:ddft_spinodal} shows snapshots from a purely evaporative
spinodal dewetting process. Panel (c) gives the evolution of mean
liquid density with time, $\langle \rho_l \rangle \equiv A^{-1}\int
\dr \rho_l(\rr,t)$, where $A$ is the area of the substrate, and panel
(f) gives the nanoparticle structure factor $S(k)\equiv
\langle\rho_n(k)^2 \rangle$, where $\rho_n(k)$ is the Fourier
transform of $\rho_n(\rr)$.  For small times, the unstable film
develops a typical spinodal labyrinthine pattern with a typical
wavelength $2\pi/k_\mathrm{max}$ (note that the symmetry
breaking in the density profiles in all our calculations is due to the addition
of a small amplitude random noise to the density profiles at time $t=0$
and no noise is added at later times). The nanoparticles concentrate where
the remaining liquid is situated. However, they are `slow' in their
reaction: when $\rho_l$ already takes values in the range 0.08 to
0.83, the nanoparticle concentration has only deviated by about 25\%
from its mean value.  The film thins rapidly forming many small
holes. The competition for space results in a fine-meshed network of
nanoparticle deposits with a much higher concentration of particles at
the network nodes -- an effect that can not be seen within the KMC
model. Because the liquid wets the nanoparticles, some liquid always
remains on the substrate. Accounting for solvent diffusion, the rate of
the dewetting process is
increased [see Fig.~\ref{fig:ddft_spinodal}(c)] and leads to a more
strongly modulated final pattern -- i.e.\ the peaks in $S(k)$ are
higher for $M_l^c >0$ than for $M_l^c=0$ [\ref{fig:ddft_spinodal}(f)].

Fig.~\ref{fig:ddft_nucleation} shows snapshots from a dewetting
process triggered by nucleation events.  The holes nucleate at several
arbitrary places due to the random initial noise in the density
profiles, and grow to form a random polygonal network of rims of
highly concentrated solution. On a very long time scale the network
coarsens into an array of drops.  The influence of liquid transport
can be seen in the final panel of Fig.\ \ref{fig:ddft_nucleation},
where we display a plot of the average density of the liquid on the
substrate as a function of (log) time. We see that 
the liquid is able to evaporate from the substrate faster
and that for times
$t/t_l \sim10^2-10^4$, the total amount of liquid on the substrate is
less when $M_l^c=2$, than when $M_l^c=0$. However, over very long
times, the drops on the substrate slowly move and `eat-up' the network
pattern. This process is faster with solvent diffusion, leading to a faster
increase of $\langle \rho_l \rangle$ at long times. Since the liquid
wets the nanoparticles, some liquid also re-condenses back onto the
substrate.

The final example in
Figs.~\ref{fig:ddft_fingers} and \ref{fig:fingers_vs_Mlc} is the
evolution of the fingering instability for a receding dewetting
front. The fingering instability is caused by a build up of the
nanoparticles at the receding front, which collects the nanoparticles due to
  their attraction to the liquid. In
Fig.~\ref{fig:ddft_fingers}(a) we see that at early times the initially
straight front shows a rather short-wave instability; about 20 short
`fingers' can be seen. However, the finger pattern coarsens rapidly to
a stationary pattern containing only about half the initial number of
fingers. Intriguingly, the mean finger number remains constant
although at the moving contact line new branches are created and old
branches merge continuously. The occurrence of this phenomenon in the
present continuum model (DDFT) is similar to results of the KMC
\cite{Vanc08}, and proves that jamming of discrete particles is not a
necessary mechanism for causing the instability. In Fig.\ \ref{fig:ddft_fingers}(f)
we show how the average number
of fingers per unit length, $\langle f \rangle$, varies as a function
of $\alpha$, the mobility coefficient of the
nanoparticles on the wet substrate, for the case when
$M_l^c=0$. We see that as $\alpha$ is decreased, the number of fingers
increases.  This increase in $\langle f \rangle$ occurs because when
the mobility of the nanoparticles is decreased 
the front `collects' more particles (less of them diffuse
further from the front). The resulting region of high concentration solution at
the front may be `dynamically unstable': As the front velocity depends
non-linearly on the amount of particles collected, any fluctuation
along the front may trigger a transverse instability.

Fig.\ \ref{fig:fingers_vs_Mlc} shows that the finger number $\langle
f \rangle$ depends non-monotonically on the mobility coefficient for
liquid diffusion, $M_l^c$. Although the overall trend is an
increase of $\langle f \rangle$ with increasing $M_l^c$, there exists
an intermediate region ($5 \lesssim M_l^c \lesssim 7$) where $\langle
f \rangle$ slightly decreases. The overall trend results from an
increase in front velocity (due to the increase in $M_l^c$) at fixed
particle diffusivity. However, we currently have no explanation for
the intermediate slight decrease.

Note also that in all cases the instability may be strongly
amplified when the particle interactions favour the clustering of
the nanoparticles (this is not the case for the parameters used
here), where the higher concentration at the receding front leads to
a local demixing of nanoparticles and liquid, that itself enforces
the deposition of a highly branched finger pattern.  In this
`demixing regime' the instability is determined by the dynamics {\it
and} the energetics of the system whereas for the case studied in
Fig.~\ref{fig:ddft_fingers} it mainly depends on the dynamics. 

We note finally, that the fingering process can be seen as a
self-optimisation of the front motion, so that the average front
velocity is kept constant by expelling particles into the fingers. A
similar effect exists for dewetting polymer films \cite{ReSh01},
where surplus liquid is expelled from the growing moving rim which
collects the dewetted polymer.  However, front instabilities found
for dewetting polymers only result in fingers without side-branches
\cite{Reit93b} or fields of droplets left behind \cite{ReSh01}.

In this Letter we have developed a versatile DDFT that is
capable of describing the pattern formation observed in evaporating
dewetting thin films of suspensions. Since our DDFT takes
account of all the basic solvent and solute transport and phase
change processes in a consistent manner, we believe that it
will be the basis for many successful future studies of the
behaviour of suspensions and solutions at interfaces.

AJA and MJR gratefully acknowledge financial support by RCUK and
EPSRC, respectively.  We acknowledge support by the EU via grant
PITN-GA-2008-214919 (MULTIFLOW).


\begin{thebibliography}{35}
\bibitem{footnote}
 We denote it a ``semi-grand'' free energy, because
  the free energy is developed in an ensemble where the liquid is treated
  grand canonically (it is coupled to a reservoir with chemical
  potential $\mu$), whereas the fixed number of nanoparticles are treated
  canonically.
\end{thebibliography}


\end{document}